# The 1st Law of Thermodynamics in Chemical Reactions

I. A. Stepanov

Latvian University, Rainis Bulv. 19, Riga, LV-1586, Latvia

**Abstract**

In the previous papers of the author it has been shown that the 1st law of thermodynamics in chemical reactions is the following one:

$$\Delta U = \Delta Q + P\Delta V + \sum_i \mu_i \Delta N_i$$

In the present paper this theory was developed and it has been shown that the 1st law of thermodynamics in chemical reactions has the following form:

$$\Delta C = -\Delta U + \Delta A$$

$$-\Delta U = \Delta Q$$

where $\Delta C$ is the change in the chemical energy, $\Delta U$ is the change in the internal energy. Internal energy is the energy of thermal motion of molecules.

## 1. Introduction

Earlier it was supposed that the 1st law of thermodynamics in chemical reactions is the following one:

$$\Delta U = \Delta Q - P\Delta V + \sum_i \mu_i \Delta N_i$$

Recently it has been shown [1-6] that it must have the following form:

$$\Delta U = \Delta Q + P\Delta V + \sum_i \mu_i \Delta N_i$$

In the present paper this theory is developed.

## 2. Theory

The internal energy of substance is its full energy. However, chemical reaction is the result of change only in the chemical energy. In substances there is the chemical energy which is potential energy and during chemical reactions turns to mechanical work, heat or electricity. Chemical energy is the energy of external electronic shells. Let's denote the change in the chemical energy $\Delta C$. Chemical energy can turn only to the energy of thermal motion of molecules $\Delta W$ and to work $\Delta A$. Then the first law of thermodynamics for chemical reactions can be written as

$$\Delta C = -\Delta W + \Delta A$$

$$-\Delta W = \Delta Q$$

Let's denote $\Delta W$ by $\Delta U$. Then the first law of thermodynamics will look like

$$\Delta C = -\Delta U + \Delta A$$

$$-\Delta U = \Delta Q$$

The key idea of this paper is that chemical reaction is due to the change only in the chemical energy. Other types of energy remain constant.

On sees that in chemical reactions the internal energy of substance is the energy of thermal motion of molecules but not its whole energy. Pay attention that the heat of chemical reaction is the change in the internal energy of the system in isochoric and isobaric cases. Earlier it was supposed that the heat of chemical reaction in isobaric case is the difference in the enthalpy.

Let's derive the second law of thermodynamics for chemical reactions. $\delta C$ is not an exact differential because it is the sum of an exact and not an exact differential. But $\delta C/T$ will be an exact differential because $dU/T+\delta A/T$ is an exact differential. Therefore, one can introduce a new function, let's call it chemical entropy $S_C$. Then

$$dS_C \geq \delta C/T = -dU/T + \delta A/T \qquad (1)$$

Let's suppose that A is the work of expansion: $\delta A = PdV$. It is work done by chemical system. Then, if P and T are constant

$$d(-U+PV-TS_C) \leq 0$$

The Gibbs energy in chemical reaction is

$$G = -U + PV - TS_C \qquad (2)$$

Correspondingly, one can show that the Helmholtz energy in chemical reactions is

$$F = -U - TS_C$$

From (1) it follows that in reversible processes

$$TdS_C = -dU + \delta A \qquad (3)$$

According to (2)

$$dG = -dU + PdV + VdP - TdS_C - S_C dT \qquad (4)$$

Introducing (3) to (4), taking into account that $\delta A = PdV$ one gets

$$dG = VdP - S_C dT$$

Whence

$$(\partial G/\partial T)_P = -S_C \qquad (5)$$

and

$$(\partial G/\partial P)_T = V$$

There are numerous articles where dependence $\Delta G°$ on T was measured, for example [7-12]. It is the following one:

$$\Delta G° = a + bT \tag{6}$$

where a and b are constants. Accuracy of (6) is very high. Eq. (6) is true in temperature intervals a few hundreds grades, $d\Delta G°/dT = b = $const. But

$$(\partial \Delta G°/\partial T)_P = -\Delta S°$$

can not be constant because

$$(\partial \Delta S/\partial T)_P = (C_{P2} - C_{P1})/T$$

where $C_{Pi}$ are heat capacities of the products and of the reactants. If $\Delta S°$ is constant then $C_{P2} = C_{P1}$ which is an absurd. Using (6), it is impossible to calculate $\Delta C_P$.

## 3. Experimental Check and Discussion

In [11] the following reaction was given:

$$CdS(sol) = Cd(liq) + S(liq), \quad T = 640\text{-}690 \text{ K}$$

$$\Delta G^0 = -164295 + 31{,}37T \pm 310 \text{ J/mol} \tag{7}$$

Eq. (7) can be written in the form

$$\Delta G^0 = a + bT + cT^2 + \delta \tag{8}$$

where $|cT^2 + \delta| \leq |\Delta|$, $\Delta$ is the absolute mistake of (7). From (8) it is easy to estimate $\Delta C_P$. For (7) $|\Delta C_P| < 0{,}9$ J/mol·K. Using [13], it is possible to calculate that $\Delta C_P = 8{,}5$ J/mol·K.

In [12] the following reaction was considered:

$$0{,}95Fe + 1/2O_2 = Fe_{0{,}95}O$$

$$1120 < T < 1320 \text{ K}$$

and the change in the Gibbs energy for it:

$$\Delta G^0 = -266358 + 63{,}10T \pm 629 \text{ J/mol}$$

One obtains from (8) that $|\Delta C_P|$<0,95 J/mol·K. Using [13], one may calculate that $\Delta C_P$=8,5 J/mol·K.

It is not surprising: the change in the entropy of chemical reaction will be $\Delta C/T$, not $\Delta Q/T$. $\Delta C$ depends on the temperature not so as $\Delta Q$ does. For chemical reactions

$$(\partial \Delta G°/\partial T)_P \neq -\Delta S°$$

hence it is another evidence that thermodynamics of chemical processes is not such as that of physical ones.

From (6) and (5) it follows that

$$(\partial \Delta G°/\partial T)_P = -\Delta S_C^0 = \text{const}$$

It is not surprising that $\Delta S_C^0$=const and $(\partial \Delta S_C^0/\partial T)_P \neq \Delta C_P/T$. The reason is that $(\partial \Delta C/\partial T)_P \neq C_P$ and $\delta C/T \neq dS$.

One can draw the following conclusions. The traditional thermodynamics is available only for description of physical processes, for description of chemical reactions another thermodynamics is necessary. Physical and chemical phenomena are qualitatively different things. The 1st and the 2nd laws of thermodynamics, Gibbs and Helmholtz energies for chemical reactions have another form than these for physical processes.